# Isotropic Dirac fermion and anomalous oscillator strength of zeroth Landau level transition

Zeping Shi[1,2,3†], Wenbin Wu[1,2,3†], Guangyi Wang[1,2,3†], Mykhaylo Ozerov[4], Jian Yuan[5,6], Wei Xia[5,6], Yuhan Du[1], Xianghao Meng[1], Xiangyu Jiang[1], Mingsen Zhou[1], Yuxi Chen[1], Hao Shen[1], Yanfeng Guo[5,6], Junhao Chu[2,7], Xiang Yuan[1,2,3,8*]

[1] State Key Laboratory of Precision Spectroscopy, East China Normal University, Shanghai 200241, China
[2] Key Laboratory of Polar Materials and Devices, Ministry of Education, School of Physics and Electronic Science, East China Normal University, Shanghai 200241, China
[3] Shanghai Center of Brain-Inspired Intelligent Materials and Devices, East China Normal University, Shanghai 200241, China
[4] National High Magnetic Field Laboratory, Florida State University, Tallahassee, Florida 32310, USA
[5] School of Physical Science and Technology, ShanghaiTech University, Shanghai 201210, China
[6] ShanghaiTech Laboratory for Topological Physics, ShanghaiTech University, Shanghai 201210, China
[7] Institute of Optoelectronics, Fudan University, Shanghai 200438, China
[8] Chongqing Institute of East China Normal University, Chongqing 401120, China

[†] These authors contributed equally to this work.
[*] Correspondence and requests for materials should be addressed to X.Y. (E-mail: xyuan@lps.ecnu.edu.cn)

**Abstract**

Dirac fermions, characterized by their linear dispersion and relativistic nature, have emerged as a prominent class of quasiparticles in condensed matter physics. While the Dirac equation, initially developed in the context of high-energy physics, provides a remarkable framework for describing the electronic properties of these materials, the inherent symmetry constraints of condensed matter often lead to deviations from the idealized paradigm. In particular, three-dimensional Dirac fermions in solids often exhibit anisotropic behavior, challenging the notion of perfect symmetry inherent in the Dirac equation. Here, we report the observation of isotropic massive Dirac fermions in LaAlSi through Landau level spectroscopy. The presence of three-dimensional massive Dirac fermions across the Fermi energy is demonstrated by quantized and semiclassical analyses of the magnetic field evolution of Landau level transitions. The isotropic topological nature, Fermi velocity, and Dirac mass are evidenced by the identical magneto-infrared response among the Faraday and three Voigt geometries. Furthermore, we observe an unusually large oscillator strength in the zeroth Landau level transition of the Dirac fermion, compared to transitions with higher indices. This phenomenon, supported by model calculations, can be attributed to the combined effects of the partial excitation of Dirac fermion and the resonant dielectric coupling with the Weyl plasma. Our work provides a strategy for realizing ideal quasiparticle excitations and their coupling effects in condensed matter systems, offering a platform for exploring relativistic physics.

The Dirac fermion, as introduced by P. A. M. Dirac in 1928, was a significant discovery in reconciling special relativity with quantum mechanics[1]. One of its profound implications was the prediction of antimatter, later confirmed experimentally[2]. This development not only advanced the understanding of particle spin but also laid the foundation for quantum field theory. The specific modifications of the Dirac equation lead to the prediction of Weyl fermions[3], massless particles with a definite chirality, and Majorana fermions[4] which could be their own antiparticles.

In condensed matter, the energy scales of collective excitation are usually lower than the rest mass of electrons which suggests a nonrelativistic treatment. Nevertheless, the interaction of electrons with the periodic potential in crystals can mimic the relativistic effects described by the Dirac equation[5,6]. Graphene is a prominent example[7], with its electrons exhibiting a linear energy-momentum relation, providing an analog to (2+1)-dimensional quantum electrodynamics and has spurred extensive research into its electronic properties.

Studying Dirac fermions in condensed matter systems offers distinct advantages, particularly in experimentally probing relativistic theory predictions such as Klein tunnelling[8]. However, a crucial distinction exists between the symmetries of free space and those encountered in crystals. The original Dirac equation is inherently isotropic in three dimensions[1], reflecting the perfect symmetry of free space. In contrast, crystals possess specific lattice structures that always break certain symmetries. This anisotropy is particularly evident in most of the three-dimensional (3D) Dirac semimetals, where the Fermi velocity (determined by the slope of the linear energy dispersion) along the out-of-plane direction is significantly lower than that along in-plane directions. This inherent discrepancy exacerbates the challenge of replicating the Dirac equation within condensed matter systems.

Here, we report the discovery of isotropic massive Dirac fermions in lanthanum aluminum silicide (LaAlSi). The relationship between Landau level transition energies and magnetic field strength substantiates the existence of 3D massive Dirac fermions across the Fermi energy. The isotropy in Fermi velocity and Dirac mass is corroborated by the identical magneto-infrared responses observed in Faraday and three Voigt geometries. Furthermore, we observe an unusually large oscillator strength in the zeroth Landau level transition of the isotropic Dirac fermion. Model calculations suggest that this can be attributed to the combined effects of the partial excitation of Dirac fermion and their dielectric coupling with the Weyl plasma. This finding offers a closer quasiparticle approximation to the behavior of fundamental particles in free space.

LaAlSi is a notable member of the $RAlX$ family ($R$ = La, Ce, Pr, Nd, Sm; $X$ = Si, Ge) which has drawn considerable attention due to the discovery of various electronic phenomena driven by the topological properties and their coupling to other collective excitation. LaAlSi crystallizes in a body-centered tetragonal structure with space group

I41md (109). Alternating layers of La, Al, and Si stack along the [001] direction, as shown in the inset of Fig. 1a, which breaks the inversion symmetry and facilitates the emergence of Weyl fermions. The X-ray diffraction (XRD) spectrum (Fig. 1a) and Raman spectrum (Fig. 1b) confirm the crystal structure, orientation and quality of the as-grown LaAlSi single crystal[9,10].

LaAlSi and LaAlGe exhibit nearly identical physical properties, both being non-magnetic Weyl semimetals with strong spin-orbit coupling (SOC). Each material hosts a total of 40 Weyl nodes, as confirmed by photoemission spectroscopy[11] and first-principles calculations[12]. Among them, three types of type-I Weyl nodes are located above the Fermi level (ranging from 60 to 130 meV), while type-II Weyl nodes are situated near the Fermi energy, enabling the study of Lorentz-violating physics[13]. The presence of multiple Weyl fermions leads to the multi-carrier feature in the quantum oscillations[14,15]. The intrinsic spin Hall effect[16], Fano resonance[10], superconductivity[17], and NMR shifts[18] are closely related to the presence of Weyl fermion. In contrast, substituting La with magnetic rare-earth elements (Ce, Pr, Nd, Sm) in the $R$Al$X$ compounds introduces a complex interplay between magnetism and Weyl fermions[19–21]. CeAl$X$ and PrAl$X$ compounds exhibit ferromagnetic or antiferromagnetic order[22–24], while NdAlSi and SmAlSi display long-wavelength helical magnetic order[25,26]. Since Weyl nodes have already been generated by inversion symmetry breaking, the magnetism allows the shift of these Weyl nodes through Zeeman-like interaction[27] or even leads to the emergence of additional Weyl fermions[28]. Anomalous Hall effect[29–32], topological Hall effect[33,34], nonlinear optical diode effect[35], topology stabilized magnetism[36], and temperature-induced quantum oscillation[37] are discovered, stemming from the interplay of strong exchange interactions[38], intrinsic Berry curvature[39], and Fermi surface nesting of Weyl fermions[40]. Therefore, while various gapless Weyl and nodal structures have been identified and extensively studied in the LaAlSi and the broader $R$Al$X$ family, our focus with infrared spectroscopy is to investigate topological structures with finite gaps and topological excitations with non-zero mass.

We investigate the single crystal LaAlSi using magneto-infrared spectroscopy on its (001) crystal plane. The measured magneto-reflectivity spectrum $R_B$ at a certain magnetic field **B** are normalized by the zero-field spectrum $R_0$. The obtained relative magneto-reflectivity $R_B/R_0$ are presented as stacked plots (constant offset) with the static magnetic field applied parallel (Faraday geometry, Fig. 1c) and perpendicular (Voigt geometry, Fig. 1d) to the [001] direction. The spectral range covers from mid- (MIR) to far-infrared (FIR), enabling the probing of possible low-energy quasiparticle excitations. Distinct peak and dip features are observed in both geometries. Those prominent peaks with lower energy exhibit significantly higher intensity. All peaks shift to higher energies with increasing magnetic field. These are characteristic features of inter-Landau-level transitions[41–44]. Notably, the magnetic field evolutions of these peaks are remarkably similar between the Faraday and Voigt geometries. These peaks are designated $T_0$, $T_1$, and $T_2$ in order of increasing energy.

The false-color plots in Fig. 2a and Fig. 2b provide a more comprehensive comparison, visually depicting the energy shift and intensity distribution as functions of the magnetic field. The presence of T$_0$ around 4 T indicates access to the quantum limit at this magnetic field. This low quantum limit field suggests that the observed electronic states are not associated with previously identified Weyl fermions, where the quantum limit occurs at much higher fields[14] (also refer to our transport results in Supplementary Section VI). For a more quantitative analysis, we extract optical transition energies by identifying apparent peak positions in raw spectra. Figure 2c displays the extracted energy as a function of the magnetic field. Several characteristics are identified: (1) The transition energy exhibits a convex rather than a linear dependence on the magnetic field; (2) All transitions extrapolate to the same non-zero energy at the zero field; (3) The energy separation becomes smaller for higher indices. These characteristics point out the presence of massive Dirac fermions across the Fermi energy, and the observed mass gap definitively rules out the gapless Weyl fermion excitations. The energy spectrum and Landau quantization can be described by a massive Dirac Hamiltonian:

$$H(\mathbf{k}) = \Delta \tau_z \sigma_0 + \hbar v_{Fx} k_x \tau_x \sigma_z + \hbar v_{Fy} k_y \tau_y \sigma_0 + \hbar v_{Fz} k_z \tau_x \sigma_x. \quad (1)$$

Here, 3D vector $\mathbf{k}$ refers to the wavevector; $\sigma_{x,y,z}$ ($\tau_{x,y,z}$) are the Pauli matrices for the spin (orbital) degrees of freedom; $\sigma_0$ represents the unit matrix; $\Delta$ and $v_{Fx,y,z}$ are the Dirac mass term and Fermi velocity, respectively; Subscript $x, y, z$ refer to the crystal axes $a, b, c$ as defined in the inset of Fig. 1a; $\hbar$ is the reduced Planck's constant. In the isotropic case, the physical properties of the Dirac fermion are determined by a single Fermi velocity $v_F$ and Dirac mass $m_D = \Delta/v_F^2$. The energy dispersion along any direction follows:

$$E(k) = \sqrt{\Delta^2 + \hbar^2 v_F^2 k^2}. \quad (2)$$

By considering an external magnetic field and performing the standard Peierls substitution in Eq. (1), we obtain the Landau level spectrum:

$$E_{\pm n} = \pm\sqrt{\Delta^2 + 2n\hbar^2 v_F^2/l_B^2}, \quad (3)$$

where $l_B = \sqrt{\hbar/eB}$ is the magnetic length; $n$ is the Landau index with "$\pm$" denoting conduction (positive energy) and valence band (negative energy). In Faraday geometry ($\mathbf{B} \parallel z$), the cyclotron plane is fixed on the $x$-$y$ plane, and thus $v_F$ in Eq. (3) is replaced with $v_{Fxy} = \sqrt{v_{Fx} v_{Fy}}$, the geometric mean of Fermi velocities along two principal axes[45]. The optical selection rules are determined by the transition matrix element and are summarized as $\Delta n = \pm 1$. Since we use an arbitrarily polarized light source, as shown in Fig. 2d, the transitions $-n \to n+1$ and $-(n+1) \to n$ are allowed and denoted as T$_n$. For a particle-hole symmetric system, these two transitions are indistinguishable in the spectrum. If strong anisotropy existed in the cyclotron plane, we would expect to observe additional selection rules[46]. The optical transition energy reads:

$$\omega_n(B) = \sqrt{\Delta^2 + 2n\hbar^2 v_F^2/l_B^2} + \sqrt{\Delta^2 + 2(n+1)\hbar^2 v_F^2/l_B^2}. \qquad (4)$$

On the basics of Eq. (4), these three characteristics can be well explained, and transition energies are well fitted as curves plotted in Fig. 2c. The Fermi velocity and Dirac mass term are determined as $v_{Fxy} = 4.743 \times 10^5$ m/s and $\Delta = 27.3$ meV, then the Dirac mass is obtained: $m_{Dxy} = 0.0214\, m_0$, where $m_0$ is the free electron mass.

Although the original Dirac equation and Dirac fermions are isotropic, Dirac fermions discovered in 3D condensed matter are generally anisotropic, of which Fermi velocities vary with the momentum directions. This anisotropy causes the Landau level spectrum to depend on the direction of the magnetic field, i.e., the plane of cyclotron motion. In the Faraday geometry, the magnetic field is parallel to the z-axis ([001] direction). Thus, the effective Fermi velocity extracted from interband-Landau-level transitions incorporates information from both the x- and y-axes. We experiment with the Voigt geometry to measure the anisotropic properties within the same sample, which overcomes the challenges of detecting different crystal planes for magneto-FIR spectroscopy. Note that $\Delta n = \pm 1$ is still valid in Voigt geometry, and we leave a more detailed discussion in Supplementary Section I. In this configuration, as depicted in the inset of Fig. 1d, the magnetic field is applied parallel to the y-axis ([010] direction), influencing the Landau level spectrum through $v_{Fzx} = \sqrt{v_{Fz} v_{Fx}}$. If in-plane/out-of-plane anisotropy is present, the corresponding Fermi velocity would differ, leading to distinct Landau level spectra between the Faraday and Voigt geometries[47]. Surprisingly, the extracted transition energies and their magnetic field evolution are almost identical in both geometries, as shown in Fig. 2c. With fixed $\Delta$, we obtain the fitted Fermi velocity $v_{Fzx} = 4.767 \times 10^5$ m/s and corresponding Dirac mass $m_{Dzx} = 0.0211\, m_0$. This finding provides strong evidence for the isotropy of the Dirac fermions in LaAlSi. Figure 2e schematically shows the isotropic massive Dirac band with identical Fermi velocities along in-plane ($v_{F\parallel}$) and out-of-plane ($v_{F\perp}$) directions.

Besides fitting with the quantized model, we further utilize a model-independent semiclassical analysis to map energy dispersion directly based on observed optical transitions. The semiclassical Lifshitz–Onsager relation specifies that the extremal area in reciprocal space is quantized as $S_n = 2\pi eB(n+\gamma)/\hbar$, corresponding to the $n$th Landau level. The phase factor $\gamma$ describes the Berry phase $\Phi_B$ associated with explored band through $\gamma = 1/2 - \Phi_B/2\pi$. For Dirac and Weyl fermions, a $\pi$ Berry phase is expected and leads to $\gamma = 0$, unless the Fermi energy surpasses the Lifshitz transition point[48]. The momentum is given by $k_n = \sqrt{2eB(n+\gamma)/\hbar}$ with the approximation of $S_n = \pi k_n^2$. Assuming the crystal momentum conservation during optical transitions and considering the optical selection rule $\Delta n = \pm 1$, for a specific interband-Landau-level transition $T_n$, we modify the momentum to $k_n = \sqrt{2eB(n+1/2+\gamma)/\hbar}$, where the additional "1/2" term originates from the Landau index shift dictated by the selection rule[49]. Thus, the joint profile $E_c - E_v$ of the associated conduction band and valence band can be derived from the curve where all measured transition energies $\omega_n$ and corresponding momenta $k_n$ collapse. As shown in Fig. 3, by setting $\gamma = 0$, extracted energies in Faraday geometry (panel (i)) and **B** ∥ *y* Voigt geometry (panel (ii)) successfully collapse on the same joint band profile of $E(k) = \pm\sqrt{\Delta^2 + \hbar^2 v_F^2 k^2}$ (blue curves). The profile parameters are given by the above-mentioned fitting of inter-Landau level transitions. This consistency further verifies the isotropic property and the index $n$ assignment.

The nearly identical Landau level spectra in the *x-y* and *y-z* planes eliminate out-of-plane anisotropy. It should be noted that verifying all directions is essential to conclude the overall isotropy and extract the absolute values of the Fermi velocity along each axis, although LaAlSi possesses a $C_4$ symmetry. Consequently, we carry out the remaining Voigt geometries in our home-built system (up to 12 T), with the magnetic field applied parallel to the *x*-axis ([100] direction). The false-color and stacked plot are provided in Supplementary Section II. As illustrated in Fig. 3a(iii), all interband-Landau-level transition energies converge to the same joint band profile (blue curve), evidencing that the Landau level spectrum is consistent with the other two geometries (See detailed comparison among all experimental geometries in Supplementary Section III). Table I summarizes the fitted Fermi velocities and corresponding Dirac mass for each cyclotron plane. Then, the Fermi velocity along each axis is further calculated and exhibited in Table II.

To verify the isotropy of Dirac fermions in LaAlSi more rigorously, we also examined the cyclotron plane where all principal axes contribute. This approach helps to confirm the rotational band isotropy rather than $C_4$ symmetry and provides a more comprehensive assessment of isotropy. We performed magneto-infrared spectroscopy in Voigt geometry with the magnetic field oriented along the [110] direction. With the presence of anisotropy, a distinct effective Fermi velocity and Landau level spectrum compared to the other two Voigt geometries would be observed, as reported in TaAs (ref. [50]). However, our results (Fig. 3a(iv)) again show consistency with the isotropic behavior observed in other geometries (blue curve). This comprehensive analysis confirms the isotropy of the massive Dirac fermions in LaAlSi, restoring inherent symmetry in the original Dirac equation.

Figure 3b exhibits all experimental transition energies (including all three transition indices from all geometries) that collapse on a single massive Dirac band scaling $\sqrt{\Delta^2 + \hbar^2 v_F^2 k^2}$. There is a finite gap at zero momentum and a nearly linear energy dispersion at large momentum that approaches the limit of zero mass (black dashed curves). The scaling of all data points aligns well with the fitting result from the quantized approach (blue curve). Importantly, the peak energies from different magneto-optical geometries (one Faraday and three Voigt geometries, denoted by different symbols) perfectly converge to the same scaling, further confirming the isotropy of the massive Dirac fermions. The analysis confirms that $\gamma = 0$, as deviation from this value would result in noticeable deviations from a single scaling curve (Fig. 3c).

We compare the isotropic characteristics of LaAlSi with 3D Dirac and Weyl materials reported in the literature. To quantify the isotropy extent of the Dirac/Weyl bands, we define an isotropy factor $I \equiv 1 - |\Delta v|/v_{F\parallel}$ where $\Delta v = v_{F\parallel} - v_{F\perp}$ is the difference between velocities along in-plane and out-of-plane directions. The more isotropic the band, the closer this factor approaches 1; conversely, the closer it approaches 0. To the best of our knowledge, we list the crystals whose Fermi velocities along different directions have been probed in experiments, including magneto-infrared spectroscopy[47,51–53], ARPES[54–59] and quantum oscillations[60–66]. In most of these materials, as depicted in Fig. 4, the out-of-plane Fermi velocity is considerably smaller than the in-plane values. While other materials exhibit isotropy factors below 0.7, LaAlSi stands out with a notably high factor of 0.997. This substantial isotropy makes LaAlSi a particularly noteworthy example of isotropic relativistic 3D fermions in condensed matter physics, which provide a Lorentz-invariant platform for exploring relativistic physics and parent materials system for investigating topological effects.

A closer examination of the magneto-infrared spectra in Figs. 1b and c reveals an unusual characteristic: the intensity of the $T_0$ peak is significantly higher compared to other peaks. Typically, in the quantum limit, the oscillator strength of interband-Landau-level transition gradually diminishes with increasing Landau index. This behavior can be described by the Kubo formula[67], where the transition matrix elements and joint density of states for different Landau level transitions are generally similar[68]. Consequently, in LaAlSi, despite the energy of $T_1$ approximately reaching twice that of $T_0$, the observed oscillator strength disparity is considerable and beyond expectation.

One possible explanation for this anomaly is the effect of partial excitation and the broadening effect of Dirac Fermions. In the case of negligible level broadening $\Gamma \sim 0$, as the schematic plot in Fig. 5a(i), only the transition of $-1 \rightarrow +0$ contributes to $T_0$, while the transition of $-0 \rightarrow +1$ is forbidden due to fully unoccupied $-0$ band bottom, i.e., the Pauli blocking effect. Note that the discussed physics in the article is particle-hole symmetric, and we set the Fermi level in the valence band to align with the carrier type determined by Hall effect measurement (Supplementary Section V). On the other hand, due to the thermal- and scattering-induced Landau levels broadening ($\Gamma \gg 0$), transitions previously forbidden for $T_0$ become partially allowed as presented in Fig. 5a(ii), significantly enhancing the intensity of $T_0$ compared with other transitions. This effect is particularly pronounced in systems accessing deep quantum limits and where the quasiparticle energy scales are relatively small, as is the case of LaAlSi here. In Fig. 5a, panels (iii) and (iv) illustrate theoretical optical conductivity corresponding to the two cases in panels (i) and (ii), respectively. A preliminary examination from peak heights shows that incorporating the broadening effect into the Kubo formula increases the intensity of $T_0$. Figure 5a(v) provides the comparison of spectral weight (SW), in which the spectral weight of corresponding $T_2$ normalizes each case. With more extensive Landau level broadening, the contribution from $-0 \rightarrow +1$ transition renders the relative spectral weight of $T_0$ significantly increased.

Another contributing factor is the presence of numerous Weyl cones and corresponding Weyl fermions in LaAlSi, many of which are near the Fermi surface[14], contributing to quasiparticle excitations known as the Weyl plasma[69]. According to an existing report[13], the plasma energy is approximately 100 meV, coinciding with the $T_0$ transition. Although the Weyl plasma does not directly influence the optical conductivity of the $T_0$ transition, it can resonantly couple with this transition in the reflectivity spectrum, so-called dielectric coupling. This phenomenon can be verified using the Drude-Lorentz function[70]. As shown in the upper panel of Fig. 5b, we set the Lorentz peak located at (pink curves) and far away from (blue curves) the plasma energy. The calculated reflectivity spectra in the lower panel of Fig. 5b reveal that, unlike a Lorentz excitation on a flat background, the presence of a strongly varying dielectric background, such as the bulk Weyl plasmon in our case, leads to pronounced peak and dip features in the reflectivity—a phenomenon known as resonant dielectric coupling. Consequently, the approximation of $T_0$ and plasma energy significantly amplifies the spectral feature on the reflectivity spectrum. Incorporating the two discussed mechanisms, we theoretically calculate the magneto-reflectivity spectra based on the Kubo formula and Drude-Lorentz function in Fig. 5c, which reproduce the experimental results. More details on calculation are provided in the Supplementary Section IV.

As revealed by our transport results and previous literature[14], a series of Weyl fermions coexist with the discovered massive Dirac fermion in LaAlSi, but do not contribute to the quantized features on magneto-infrared spectra. This absence may be attributed to (1) the relatively short quasiparticle lifetime within this specific material system, (2) the high Fermi level with respect to the Weyl point leading to less pronounced quantized level spacing, as evidenced by the high frequency of quantum oscillations (Supplementary Section VI), (3) strong scattering intensity and weak oscillator strength of high-index interband-Landau-level transitions, (4) influence from other bands existing near the Fermi level. Notably, these Weyl fermions exhibit anisotropic characteristics (especially for type-II Weyl fermions), suggesting they may contribute to our spectra in a non-quantized manner, which results in different background characteristics across various configurations (see Figs. 1c and d). Nevertheless, the conclusion regarding the isotropic nature of the massive Dirac fermions is based on quantized Landau level spectroscopy and remains unaffected by these background variations.

In conclusion, our study presents compelling evidence of isotropic massive Dirac fermions in LaAlSi. The identification of these fermions is substantiated by both quantized and semiclassical analyses on Landau level spectroscopy. We demonstrate that the isotropy of the topological nature, Fermi velocity, and Dirac mass is preserved across the three-dimensional crystal structure, as evidenced by consistent magneto-infrared responses in various experimental geometries. The observed enhancement in oscillator strength of the zeroth Landau level transition, compared to higher-index transitions, is attributed to the joint effect of partial excitation of Dirac fermions and resonant dielectric coupling with Weyl plasma. The observed isotropy of the 3D Dirac fermions, combined with the enhanced oscillator strength of the zeroth Landau level transition, provides a close approximation of Dirac quasiparticles and their interplay with Weyl fermions in condensed matter systems. This work offers a direct emulation of fundamental particles in free space and provides new insights into studying relativistic physics in solid-state environments.

## Methods
### Single-crystal growth
The single crystals of LaAlSi were synthesized using the standard self-flux method as previously reported[14]. Starting materials of La, Al and Si (molar ratio of 1:10:1) were mixed and placed into an alumina crucible. The excess Al plays the role of the fluxing agent. The crucible was sealed into a vacuum quartz ampoule and then heated to 1100 °C within 10 h. This temperature was held for 20 h, aiming for a sufficient reaction. After slowly cooling to 750 °C at a rate of 0.5 °C/h, the ampoule was quickly transferred to a high-speed centrifuge. With liquated Al removed, millimeter-size LaAlSi crystals with shiny surfaces were selected in the alumina crucible.

### Crystal characterization
The crystal orientation was determined by the XRD analysis (PANalytical Empyrean). We utilize the home-built Raman spectroscopy system to examine the Raman-active phonon modes with 632.8nm He-Ne laser excitation at room temperature. Magneto-transport measurements were conducted using a superconducting magnet with the standard lock-in technique.

### Magneto-infrared measurement up to 17.5 T
Magneto-infrared spectra up to 17.5 T were collected at SCM-3 of the National High Magnetic Field Laboratory (NHMFL). The LaAlSi crystal was placed into a 17.5 T liquid-helium-cooled superconducting magnet and cooled to ~5 K. The collimated infrared beam from Fourier-transform infrared (FTIR) spectrometer (Bruker 80V) was guided through a brass light pipe in vacuum and focused on the (001) plane. Static magnetic fields were applied along [001] and [010] directions for Faraday and Voigt geometries, respectively. The reflected infrared beam was collected by the nearby bolometer, and then spectra were obtained by Fourier analysis.

### Magneto-infrared measurement up to 12 T
Magneto-infrared spectra up to 12 T were collected in the home-built magneto-infrared spectroscopy system. The LaAlSi crystal was placed into the variable temperature insert (VTI) of a 12 T closed-cycle superconducting magnet (Oxford Instruments TeslatronPT) and cooled to ~ 5 K. The collimated infrared beam from the FTIR spectrometer (Bruker 80V) was guided into VTI and focused on the (001) plane of the sample by a 90° off-axis parabolic mirror. The static magnetic fields were applied along [100] and [110] directions for two Voigt geometry experiments. With the guidance of a brass light pipe, the reflected infrared beam propagated out of the VTI and was collected by a liquid-nitrogen-cooled HgCdTe detector, and then spectra were obtained by Fourier analysis.

**Data availability**
All data that support the findings of this study are available from the corresponding author upon request.

**Competing interests**
The authors declare no competing interests.

**Figures and Tables**

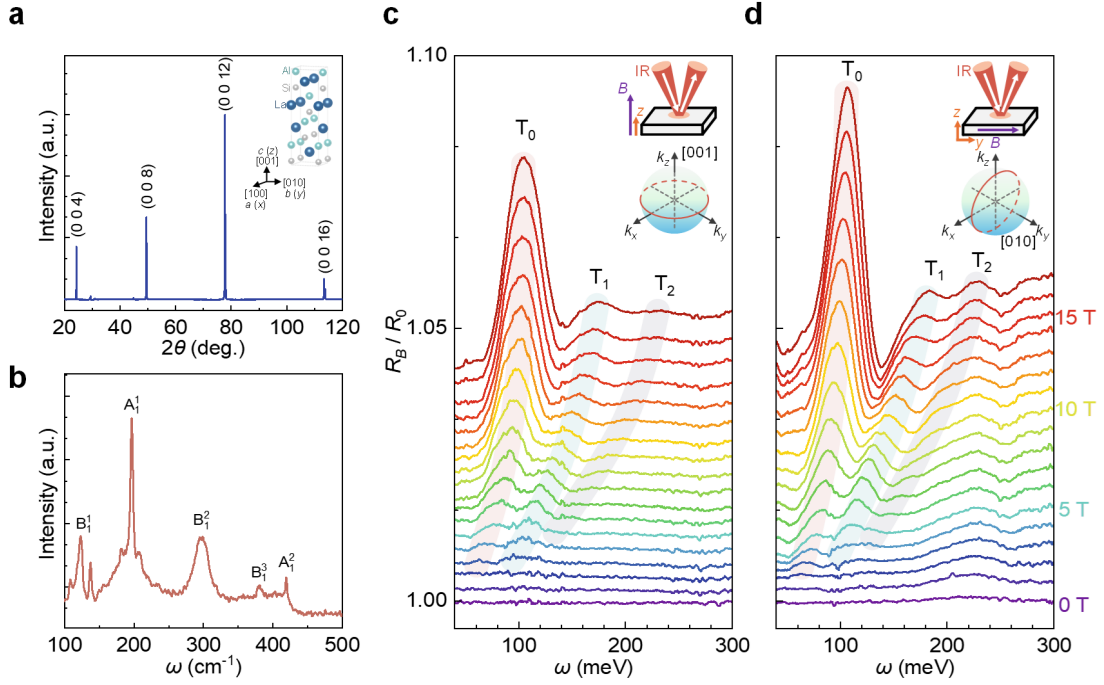

**Fig. 1 | Magneto-infrared spectroscopy of LaAlSi. a,** Single-crystal X-ray diffraction pattern, identifying the (001) plane of LaAlSi. Inset: body-centered tetragonal structure of LaAlSi. **b,** Raman spectrum of single crystal LaAlSi. **c** and **d,** Stacked plots of relative magneto-reflectivity $R_B/R_0$ spectra measured in Faraday ($\mathbf{B} \parallel z$) and Voigt ($\mathbf{B} \parallel y$) geometries, respectively, with the infrared (IR) beam focused on the (001) plane. Three series of peak features, denoted by $T_0$, $T_1$, and $T_2$, appear and evolve with the increasing magnetic fields, which are traced by red, green, and grey curves, respectively. Insets: schematic plots of magneto-optic geometries and cyclotron planes (red circle).

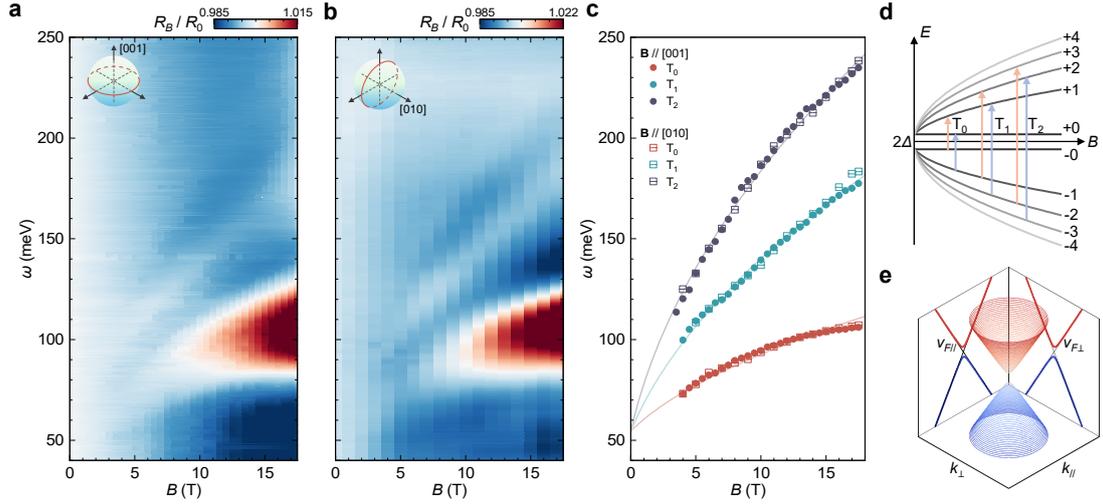

**Fig. 2 | Interband-Landau-level transitions and isotropic massive Dirac fermions.**
**a** and **b**, False-color plots of relative magneto-reflectivity $R_B/R_0$ of LaAlSi measured in Faraday ($\mathbf{B} \parallel z$) and Voigt ($\mathbf{B} \parallel y$) geometries, respectively. Insets reveal the corresponding cyclotron planes. **c,** Extracted energies of interband-Landau-level transitions from (**a**) and (**b**) denoted by dots and center-horizontal-line squares, respectively. Different colors are utilized to distinguish transitions of $T_0$ (red), $T_1$ (green) and $T_2$ (grey). Transitions energy from two geometries overlap and converge to the same fitting curves from the Landau level spectrum of massive Dirac fermion (**d**). The orange (blue) arrows in (**d**) denote the optical selection rule of $\Delta n = +1$ ($\Delta n = -1$). **e,** Schematic plot of isotropic massive Dirac band and its cross-section.

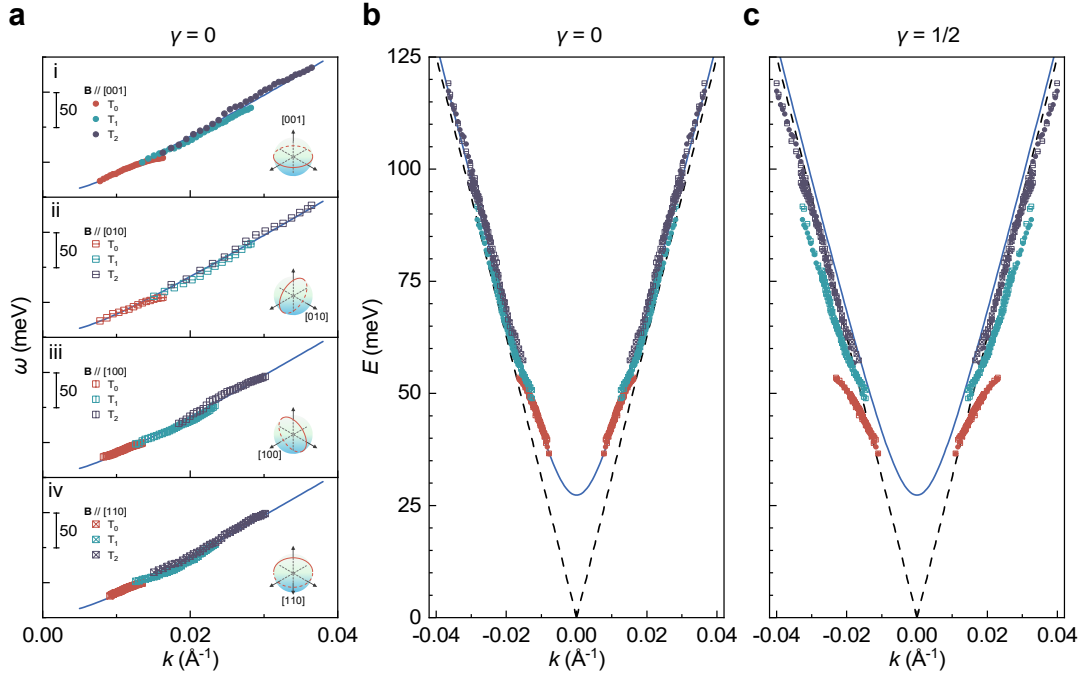

**Fig. 3 | Semiclassical analysis of spectra and addressing the shape of isotropic massive Dirac band. a,** Transition energies $\omega$ scaling with momenta given by the semiclassical Lifshitz-Onsager relation at different magneto-optical geometries: Faraday geometry with **B** ∥ [001] (i); Voigt geometries with **B** ∥ [010] (ii), **B** ∥ [100] (iii) and **B** ∥ [110] (iv). With nontrivial Berry phase $\pi$, all extracted transition energies collapse to the same joint massive Dirac band profile (blue curve in each panel), indicating an isotropic massive Dirac band in LaAlSi. **b** and **c,** A direct comparison between the cases of $\gamma = 0$ and $\gamma = 1/2$. The obtained massive Dirac band (blue curve) $E = \sqrt{\Delta^2 + \hbar^2 v_F^2 k^2}$ exhibits a finite gap (Dirac mass) at zero-momentum and a nearly linear energy dispersion in high energy approaching the zero-mass limit (black dashed curves).

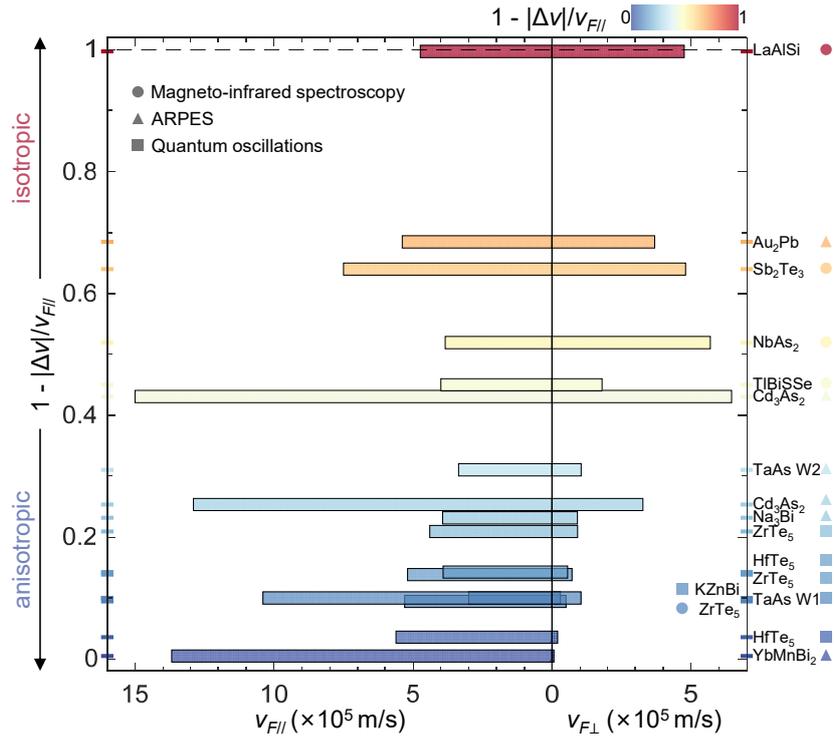

**Fig. 4 | A comparison of band isotropy among various 3D Dirac/Weyl materials.** A factor $I = 1 - |\Delta v|/v_\parallel$ is defined to describe the isotropy extent of 3D Dirac/Weyl materials. The more isotropic the band, the closer this factor approaches 1; conversely, the closer it approaches 0. The crystals are listed where Fermi velocities along different directions have been probed in experiments. The isotropy factor of LaAlSi reaches 0.997.

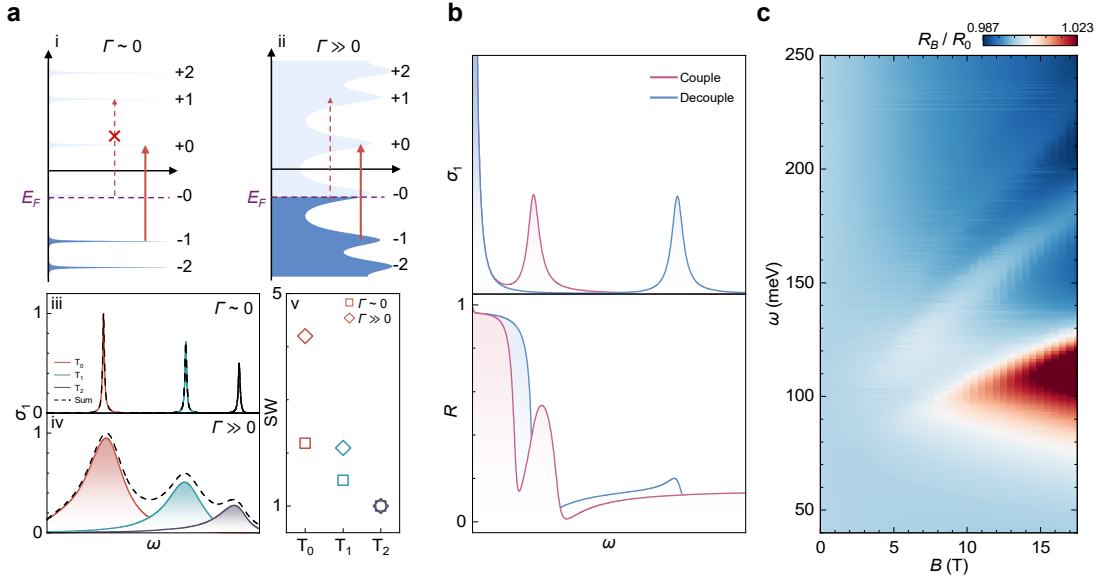

**Fig. 5 | Theorical reproduction of magneto-infrared spectra of LaAlSi. a,** Comparison of optical conductivity in different extents of Landau level broadening. (i) and (ii): the schematic drawings of Landau levels with negligible ($\Gamma \sim 0$) and considerable ($\Gamma \gg 0$) energy broadening, respectively. In a quantum-limit system, $T_0$ is intrinsically contributed by the transition of $-1 \rightarrow +0$ (red solid arrow), and the transition of $-0 \rightarrow +1$ (red dashed arrow) is generally forbidden because of the Pauli blocking effect. The energy broadening of $-0$ Landau level and thermal excitation can partially allow $-0 \rightarrow +1$. Calculated from optical conductivity spectra (iii) and (iv), the spectral weight (v) of $T_0$ is significantly enhanced with an increased extent of broadening. **b,** Simulation of spectral feature amplified by the bulk plasma. In the upper panel, two optical conductivity spectra possess identical Drude components, but the Lorentz components with the same spectral weight are located at (pink curve) and far away from (blue curve) the plasma energy. Lower panel: corresponding reflectivity spectra, indicating that the dielectric coupling with the plasma edge can significantly amplify the spectral feature of the Lorentz component. **c,** Theoretical magneto-infrared spectra calculated based on the Kubo formula and Drude-Lorentz function.

## Table 1 | Fermi velocity and Dirac mass extracted from different experimental geometries.

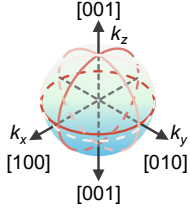

| Geometry | $v_F$ ($10^5$ m/s) | $m_D$ ($m_0$) |
|---|---|---|
| **B ∥ [001]** | 4.743 | 0.02136 |
| **B ∥ [010]** | 4.767 | 0.02114 |
| **B ∥ [100]** | 4.733 | 0.02145 |
| **B ∥ [110]** | 4.773 | 0.02109 |

## Table 2 | Fermi velocities along different directions

| $v_{Fx}$ ($10^5$ m/s) | $v_{Fy}$ ($10^5$ m/s) | $v_{Fz}$ ($10^5$ m/s) |
|---|---|---|
| 4.778 | 4.709 | 4.756 |